\documentclass[final,3p]{elsarticle}

\usepackage{amssymb}
\usepackage{graphicx}
\usepackage{array}
\usepackage{mdwmath}
\usepackage{mdwtab}
\usepackage{eqparbox}
\usepackage{subfig}
\usepackage{amsmath}
\usepackage[numbers]{natbib}

\DeclareMathOperator*{\argmin}{arg\,min}

\DeclareMathOperator{\oD}{D}                                              
\DeclareMathOperator{\oH}{H}                                              
\newcommand{\cR}{\mathcal{R}}                                          

\journal{Computer Standards \& Interfaces}

\begin{document}

\begin{frontmatter}

\title{On Content-Based Recommendation and\\User Privacy in Social-Tagging Systems}
\author{Silvia Puglisi}
\ead{silvia.puglisi@upc.edu}
\author{Javier~Parra-Arnau}
\ead{javier.parra@entel.upc.edu}
\author{Jordi Forn\'e\corref{ttl:CorrespAuthor}}
\ead{jforne@entel.upc.edu}
\author{David~Rebollo-Monedero}
\ead{david.rebollo@entel.upc.edu}

\cortext[ttl:CorrespAuthor]{Corresponding author. Tel.: +34 93 401 1871.}
\address{Department of Telematics Engineering, Universitat Polit\`ecnica de Catalunya (UPC)\\%
C.\ Jordi Girona 1-3, 08034 Barcelona, Spain}

\begin{abstract}
Recommendation systems and content filtering approaches based on annotations and ratings, essentially rely on users expressing their preferences and interests through their actions, in order to provide personalised content. This activity, in which users engage collectively has been named social tagging, and it is one of the most popular in which users engage online, and although it has opened new possibilities for application interoperability on the semantic web, it is also posing new privacy threats. It, in fact, consists of describing online or offline resources by using free-text labels (i.e. tags), therefore exposing the user profile and activity to privacy attacks. Users, as a result, may wish to adopt a privacy-enhancing strategy in order not to reveal their interests completely. Tag forgery is a privacy enhancing technology consisting of generating tags for categories or resources that do not reflect the user's actual preferences. By modifying their profile, tag forgery may have a negative impact on the quality of the recommendation system, thus protecting user privacy to a certain extent but at the expenses of utility loss. The impact of tag forgery on content-based recommendation is, therefore, investigated in a real-world application scenario where different forgery strategies are evaluated, and the consequent loss in utility is measured and compared.
\end{abstract}

\begin{keyword}
Information Privacy; Privacy-enhancing technology; Privacy Risk; Recommendation system; Social Web; Collaborative tagging; Tag forgery.
\end{keyword}

\end{frontmatter}


\section{Introduction}
\label{introduction}
\noindent
{R}{ecommendation} and information filtering systems have been developed to predict users' preferences, and eventually use the resulting predictions for a variety of services, from search engines to resources suggestions and advertisement. The system functionality relies on users implicitly or explicitly revealing their activity and personal preferences, which are ultimately used to generate personalised recommendations.

Such annotation activity has been called \emph{social tagging} and it consists of users collectively assigning keywords (i.e.\ \emph{tags}) to real life objects and web-based resources that they find interesting. Social tagging is currently one of the most popular online activities. Therefore, different functionalities have been implemented in various online services, such as Twitter, Facebook, YouTube, and Instagram, to encourage their users to tag resources collectively.

Tagging involves classifying resources according to one's experience. Unlike traditional methods where classification happens by choosing labels from a controlled vocabulary, in social tagging systems users freely choose and combine terms. This is usually referred to as free-form tag annotation, and the resulting emergent information organisation has been called \emph{folksonomy}.

This scenario has opened new possibilities for semantic interoperability in web applications. Tags, in fact, allow autonomous agents to categorise web resources easily, obtaining some form of semantic representation of their content.
However, annotating online resources poses potential privacy risks, since users reveal their preferences, interests and activities. They may then wish to adopt privacy-enhancing strategies, masquerading their real interests to a certain extent, by applying tags to categories or resources that do not reflect their actual preferences. Specifically, \emph{Tag forgery} is a privacy enhancing technology (PET) designed to protect user privacy, by creating bogus tags in order to disguise real user's interests. As a perturbation-based mechanism, tag forgery poses an inherent trade-off between privacy and usability. Users are able to obtain a high level of protection by increasing their forgery activity, but this can substantially affect the quality of the recommendation.

The primary goal of this work is to investigate the effects of tag forgery to content-based recommendation in a real-world application scenario, studying the interplay between the degree of privacy and the potential degradation of the quality of the recommendation. An experimental evaluation is performed on a dataset extracted from Delicious~\cite{a20}, a social bookmarking platform for web resources. In particular, three different tag forgery strategies have been evaluated, namely: \emph{optimised tag forgery}~\cite{Rebollo10IT}, \emph{uniform tag forgery} and \emph{TrackMeNot} (TMN)~\cite{Howe06B}, the last consists of simulating a possible TMN like agent, periodically issuing randomised tags according to popular categories.

Using the dataset and a measure of utility for the recommendation system, a threefold experiment is conducted to evaluate how the application of tag forgery may affect the quality of the recommender. Hence, we simulate a scenario in which users only apply one of the different tag forgery strategies considered. Measures of the recommender performances are computed before and after the application of each PET, obtaining an experimental study of the compromise between privacy and utility.

To the best of our knowledge, this is the first systematic evaluation of the impact of applying perturbation-based privacy technologies on the usability of content-based recommendation systems. For this evaluation, both suitable privacy and usability metrics are required. In particular, as suggested by Parra et al.~\cite{Parra13FGCS}, the KL divergence is used as privacy metric of the user profile; while the quality of the recommendation is computed following the methodology proposed by Cantador el al. ~\cite{a02}.

This paper is organised as follows: Section~\ref{SOA} introduces the state of the art. Section~\ref{adv model} describes the adversary model considered. Section~\ref{arch} explains a possible practical application of the proposed PET through the implementation of a communication module. Section~\ref{eval} discusses the evaluation methodology and obtained results. Section~\ref{conclusions} presents the conclusions that can be derived from the presented results, while also introducing future research lines.

\section{State of The Art}
\label{SOA}
\noindent
In recommendation systems employing tags or in any system allowing resource annotation, users decide to disclose personal data in order to receive, in exchange, a certain benefit. This earned value can be quantified in terms of the customised experience of a certain product~\cite{a01}.For such a recommendation system to work, and successfully propose items of interest, user preferences need to be revealed and made accessible partially or in full, and thus exposed to possible privacy attacks.

When a user expresses and shares their interests by annotating a set of items, these resources and their categorisation will be part of their activity. The recorded users' activities will allow the used platform to ``know more" about each of them, and therefore suggesting over time useful resources. These could be items similar to others tagged in the past, or simply close to the set of preferences expressed in their profile. In order to protect their privacy, a user could refrain from expressing their preferences altogether. While in this case an attacker would not be able to build a profile of the user in question, it would also become impossible for the service provider to deliver a personalised experience: the user would then achieve the maximum level of privacy protection, but also the worst level of utility.

Various and numerous approaches have been proposed to protect user privacy by also preserving the recommendation utility in the context of social tagging platform. These approaches can be grouped around four main strategies~\cite{Shen07SIGIR}: encryption-based methods, approaches based on trusted third parties (TTPs), collaborative mechanisms and data-perturbative techniques. In traditional approaches to privacy, users or application designers decide whether certain sensitive information is to be disclosed or not. While the unavailability of this data, traditionally attained by means of access control or encryption, produces the highest level of privacy, it would also limit access to particular content or functionalities. This would be the case of a user freely annotating items on a social tagging platform. By adopting traditional PETs, the profile of this user could be made available only to the service providers, but kept completely or partially hidden from their network of social connections on the platform. This approach would indeed limit the chances of an attacker profiling the user, but would, unfortunately, prevent them from receiving content suggested by their community.

A conceptually simple approach to protecting user privacy consists in a TTP acting as an intermediary or \emph{anonymiser} between the user and an untrusted information system. In this scenario, the system cannot know the user ID, but merely the identity of the TTP involved in the communication. Alternatively, the TTP may act as a \emph{pseudonymiser} by supplying a pseudonym ID' to the service provider, but only the TTP knows the correspondence between the pseudonym ID' and the actual user ID. In online social networks, the use of either approach would not be entirely feasible as users of these networks are required to authenticate to login. Although the adoption of TTPs in the manner described must, therefore, be ruled out, the users could provide a pseudonym at the sign-up process. In this regard, some sites have started offering social-networking services where users are not required to reveal their real identifiers. Social Number~\cite{SocialNumber} is an example of such networks, where users must choose a unique number as their ID.

Unfortunately, none of these approaches effectively prevents an attacker from profiling a user based on the annotated items content, and ultimately inferring their real identity. This could be accomplished in the case of a user posting related content across different platforms, making them vulnerable to techniques based on the ideas of reidentification. As an example, suppose that an observer has access to certain behavioural patterns of online activity associated with a user, who occasionally discloses their ID, possibly during interactions not involving sensitive data. The same user could attempt to hide under a pseudonym ID' to exchange information of confidential nature. Nevertheless, if the user exhibited similar behavioural patterns, the unlinkability between ID and ID' could be compromised through the exploitable similarity between these patterns. In this case, any past profiling inferences carried out by the pseudonym ID' would be linked to the actual user ID.

A particularly rich group of PETs resort to users collaborating to protect their privacy. One of the most popular is \emph{Crowds}~\cite{Reiter98ISS}, which assumes that a set of users wanting to browse the Web may collaborate to submit their requests. Precisely, a user wishing to send a request to a Web server selects first a member of the group at random, and then forwards the request to them. When this member receives the request, it flips a biased coin to determine whether to forward this request to another member or to submit it directly to the Web server. This process is repeated until the request is finally relayed to the intended destination. As a result of this probabilistic protocol, the Web server and any of the members forwarding the request cannot ascertain the identity of the actual sender, that is, the member who initiated the request.

We consider collaborative protocols~\cite{Domingo09DKE,Rebollo09COMCOM,Domingo12INS} like Crowds, not suitable for the application addressed in this work although they may be effective in applications such as information retrieval and Web search. The main reason is that users are required to be logged into online social tagging platforms. That is, users participating in a collaborative protocol would need the credentials of their peers to log in, and post on their behalf, which in practice would be unacceptable. Besides, even if users were willing to share their credentials, this would not entirely avoid profiling based on the observation of the resources annotated.

In the case of perturbative methods for recommendation systems, \cite{Polat03SDM}~proposes that users add random values to their ratings and then submit these perturbed ratings to the recommender. When the system has received these ratings, it executes an algorithm and sends the users some information that allows them to compute the final prediction themselves. When the number of participating users is sufficiently large, the authors find that user privacy is protected to some degree, and the system reaches an acceptable level of accuracy. However, even though a user may disguise all their ratings, merely showing interest in an individual item may be just as revealing as the score assigned to that item. For instance, a user rating a book called ``How to Overcome Depression'' indicates a clear interest in depression, regardless of the score assigned to this book. Apart from this critique, other works~\cite{Kargupta03ICDM,Huang05SIGMOD} stress that the use of certain \emph{randomised} data-distortion techniques might not be able to preserve privacy completely in the long run.

In line with these two latter works, \cite{Polat05SAC}~applies the same perturbative technique to collaborative filtering algorithms based on singular-value decomposition, focusing on the impact that their technique has on privacy. For this purpose, they use the privacy metric proposed by Agrawal, and Aggarwal, ~\cite{Agrawal01SIGMOD}, effectively a normalized version of the mutual information between the original and the perturbed data, and conduct some experiments with data sets from Movielens~\cite{Movielens} and Jester~\cite{Jester}. The results show the trade-off curve between accuracy in recommendations and privacy. In particular, they measure accuracy as the mean
absolute error between the predicted values from the original ratings and the predictions obtained from the perturbed ratings.

The approach considered in this study follows the idea of perturbing the information implicitly or explicitly disclosed by the user. It, therefore, represents a possible alternative to hinder an attacker in their efforts to profile their activity precisely, when using a personalised service. The submission of false user data, together with genuine data, is an illustrative example of data-perturbative mechanism. In the context of information retrieval, query forgery~\cite{Rebollo10IT} prevents privacy attackers from profiling users accurately based on the \emph{content} of queries, without having to trust the service provider or the network operator, but obviously at the cost of traffic overhead. In this kind of mechanisms, the perturbation itself typically takes place on the user side. This means that users do not need to trust any external entity such as the recommender, the ISP or their neighbouring peers. Naturally, this does not signify that data perturbation cannot be used in combination with other third-party based approaches or mechanisms relying on user collaboration.

Certainly, the distortion of user profiles for privacy protection may be done not only by means of the insertion of false activity, but also by suppression. An example of this latter kind of data perturbation may be found in Parra et al.~\cite{Parra10TB}, where the authors propose the elimination of tags as a privacy-enhancing strategy in the context of the semantic Web. This strategy allows users to preserve their privacy to a certain degree, but it comes at the cost of a degradation in the semantic functionality of the Web. Precisely, Parra et al. \cite{Parra12DKE}~investigates mathematically the privacy-utility trade-off posed by the suppression of tags, measuring privacy as the Shannon entropy of the perturbed profile, and utility as the percentage of tags users are willing to eliminate. Closely related to this work is also another study of Parra et al.~\cite{Parra12TKDE}, where the impact of tag suppression is assessed experimentally in the context of a parental control application, in terms of percentages regarding missing tags on resources on the one hand, and in terms of false positives and negatives on the other.

\section{Adversary Model}
\label{adv model}
\noindent
Users tagging online and offline resources generate what is has been called a folksonomy, that is, a set composed by all the users that have expressed at least a tag, the tags that have been used and the items that have been described through them.
Formally, a folksonomy $\mathcal{F}$ can be defined as a tuple $\mathcal{F}=\{\mathcal{T},\mathcal{U},\mathcal{I},\mathcal{A}\}$, where $\mathcal{T}=\{t_1,\ldots,t_L\}$ is the set of tags, or more generally tag categories, which comprise the vocabulary expressed by the folksonomy; $\mathcal{U}=\{u_1,\ldots,u_M\}$ is the set of users that have expressed at least a tag; $\mathcal{I}=\{i_1,\ldots,i_N\}$ is the set of items that have been tagged; and $\mathcal{A}=\{(u_m, t_l, i_n) \in \mathcal{U} \times \mathcal{T} \times \mathcal{I} \}$ is the set of annotations of each tag category $t_l$ to an item $i_n$ by a user $u_m$~\cite{a02}.

As we shall see in Section~\ref{mod profiles}, our user-profile model will rely on categorising tags into categories of interest. This will provide a certain mathematical tractability of the user profile while at the same time allowing for a classification of the user interests into macro semantic topics.

In our scenario; users assign tags to online resources, according to their preferences, taste or needs. It follows that while the user is contributing to categorise a specific content through their tags, hence adding semantic information to the whole folksonomy, their activity is revealing something regarding their interests, reducing their privacy overall.

We assume that the set of potential privacy attackers includes any entity capable of capturing the information users convey to a social tagging platform. Accordingly, both service providers and network operators are deemed potential attackers. However, since tags are often publicly available to other users of the tagging platform, any entity able to collect this information is also taken into consideration in our adversary model.

In our model, we suppose that the privacy attacker aims at profiling users through their expressed preferences, specifically on the basis of the tags posted. Throughout this work, we shall consider that the objective of this profiling activity is to \emph{individuate} users, meaning that the attacker wishes to find users whose preferences significantly diverge from the interests of the whole population of users. This assumption is in line with other works in the literature~\cite{Parra13FGCS,Parra13PhD,Parra14Entropy}.

\subsection{Modelling the User/Item Profiles}
\label{mod profiles}
\noindent
A tractable model of the user profile as a probability mass function (PMF) is proposed in~\cite{Parra10TB,Parra12DKE,Parra12TKDE,Parra13PhD} to express how each tag contributes to expose how many times the user has expressed a preference toward a specific category of interest. This model follows the intuitive assumption that a particular category is weighted according to the number of times this has been used in the user or item profile.

Exactly as in those works, we define the profile of a user $u_m$ as the PMF $p_m = (p_{m,1},\ldots, p_{m,L})$, conceptually a histogram of relative frequencies of tags across the set of tag categories $\mathcal{T}$. More formally, in terms of the notation introduced at the beginning of Section~\ref{adv model}, the $l$-th component of such profile is defined as
$$p_{m,l} = \frac{|\{(u_m,t_l,i)\in\mathcal{A} | i\in\mathcal{I}\}|}{|\{(u_m,t,i)\in\mathcal{A} | t\in\mathcal{T}, i\in\mathcal{I}\}|}.$$

\begin{figure}[tb!]
\centering\hspace*{\fill}
\subfloat[Example of user profile expressed as a PMF across a set of tag categories.]%
{\includegraphics[scale=0.25]{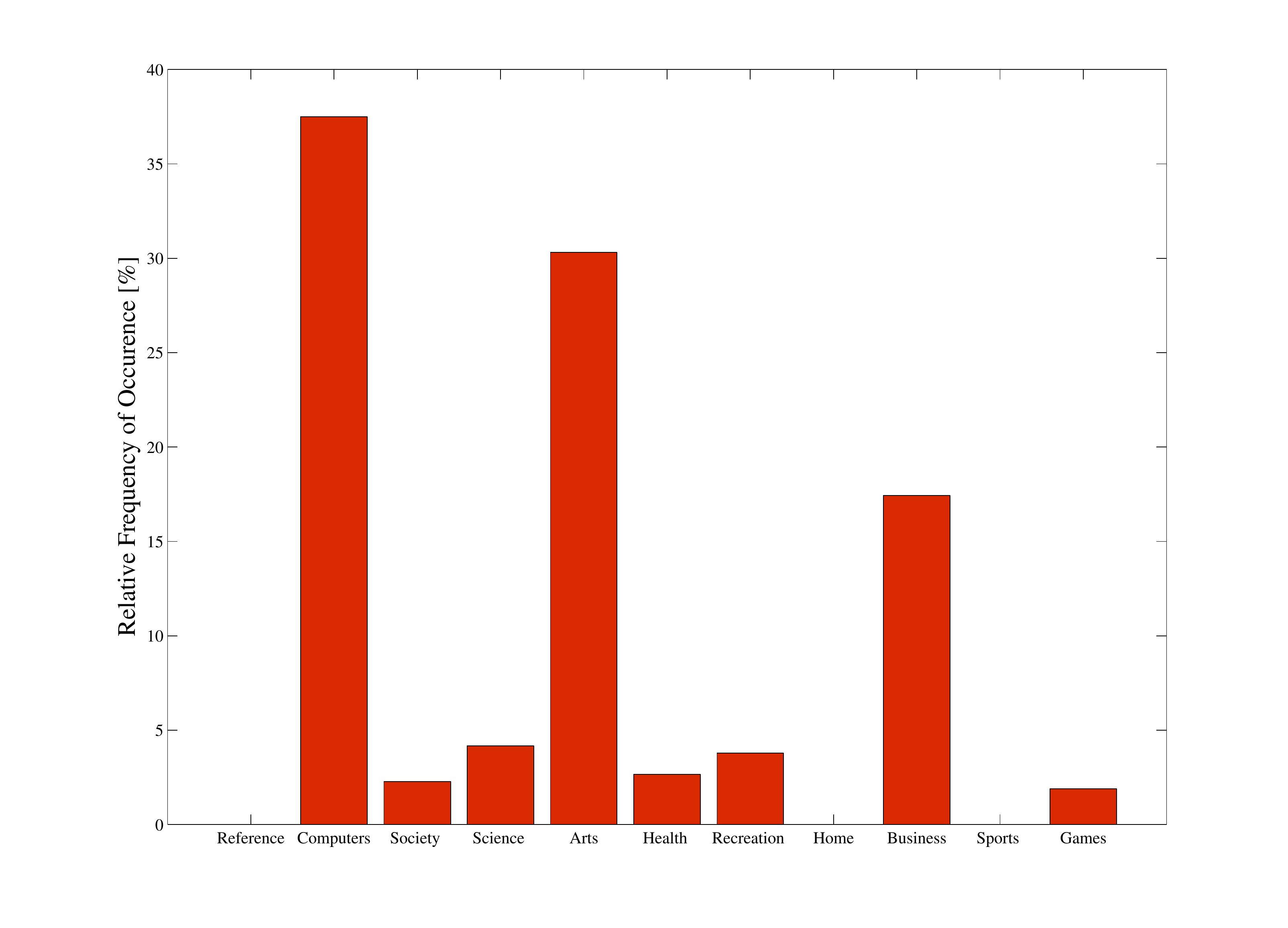}%
\label{UserProfile}}\hfill
\subfloat[Profile of the whole population of users in our dataset.]%
{\includegraphics[scale=0.25]{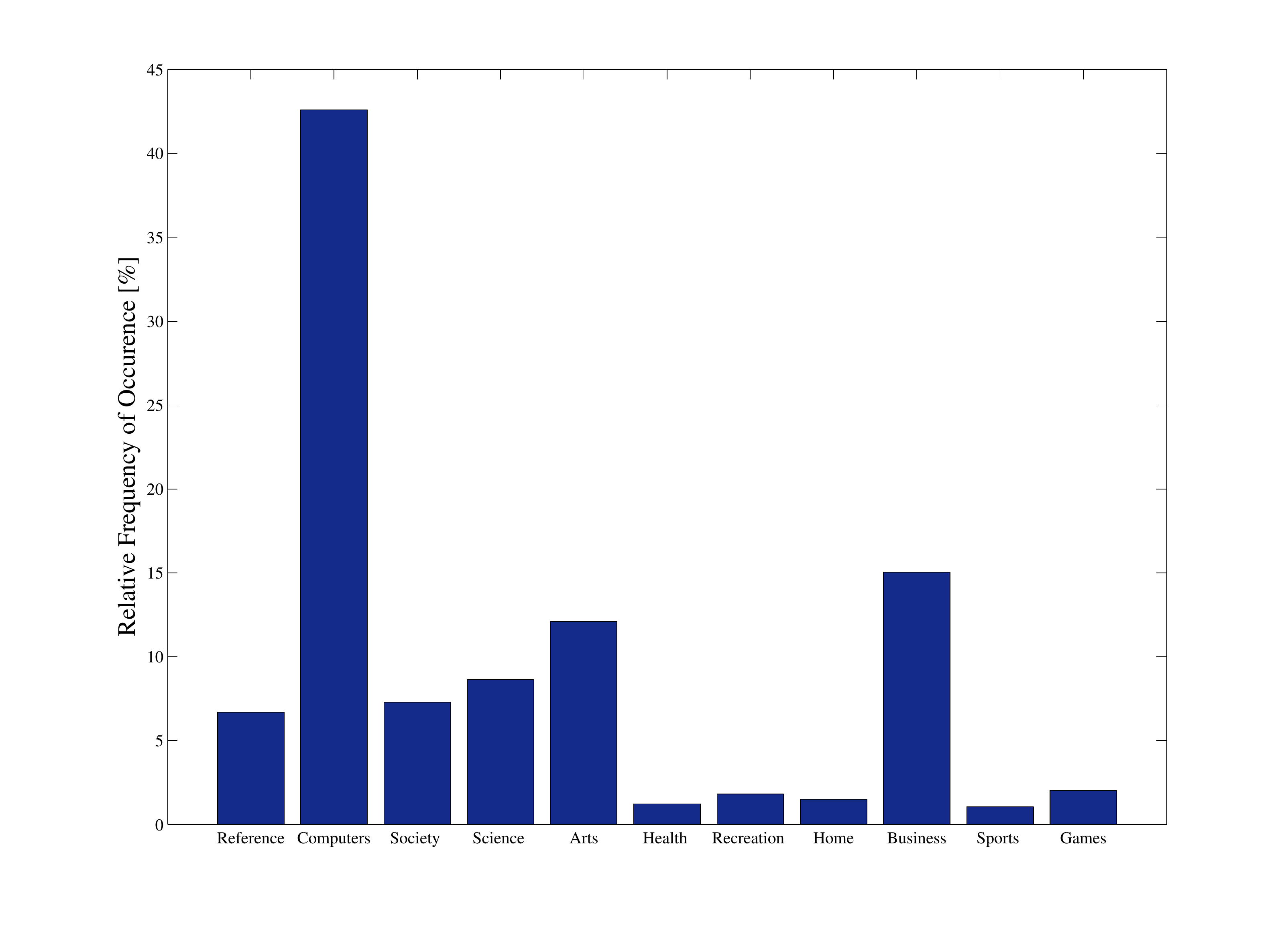}%
\label{PopProfile}}\hspace*{\fill}
\caption{We model user and item profiles as normalised histograms of tags across a set of predefined categories of interest.}
\label{PMF}
\end{figure}

Similarly, we define the profile of an item $i_n$ as the PMF $q_n =(q_{n,1},\ldots, q_{n,L})$, where $q_{n,l}$ is the percentage of tags belonging to the category $l$ which have been assigned to this item. Both user and item profiles can then be seen as normalised histograms of tags across categories of interest. Our profile model is in this extent equivalent to the tag clouds that numerous collaborative tagging services use to visualise which tags are being posted, collaboratively or individually by each user. A tag cloud, similarly to a histogram, is a visual representation in which tags are weighted according to their relevance. Figure~\ref{UserProfile} shows an example of user profile.

In view of the assumptions described in the previous section, our privacy attacker boils down to an entity that aims to profile users by representing their interests in the form of normalised histograms, on the basis of a given categorisation.
To achieve this objective, the attacker exploits the tags that users communicate to social tagging systems. This work assumes that users are willing to submit false tags, to mitigate the risk of profiling. In doing so, users gain some privacy, although at the cost of certain loss in usability. As a result of this, the attacker observes a perturbed version of the genuine user profile, also in the form of a relative histogram, which does not reflect the actual interests of the user. In short, the attacker believes that the observed behaviour characterises the actual user's profile.

Thereafter, we shall refer to these two profiles as the \emph{actual} user profile $p$ and the \emph{apparent} user profile~$t$.

\subsection{Privacy Metric}
\noindent
In this section, we propose and justify an information-theoretic quantity as a measure of user privacy in social tagging systems.
For the readers not familiar with information theory, next we briefly review two key concepts.

Recall~\cite{Cover06B} that Shannon's entropy $\oH(p)$ of a discrete random variable (r.v.) with PMF $p=(p_i)_{i=1}^L$ on the alphabet $\{1,\ldots,L\}$ is a measure of the uncertainty of the outcome of this r.v., defined as
\begin{equation*}
\oH(p)=-\sum p_i   \log p_i.
\end{equation*}
Given two probability distributions $p$ and $q$ over the same alphabet, the Kullback-Leibler (KL) divergence is defined as
\begin{equation*}
\oD(p\,\|\,q)=\sum p_i   \log \frac{p_i}{q_i}.
\end{equation*}
The KL divergence is often referred to as \emph{relative entropy}, as it may be regarded as a generalisation of the Shannon entropy of a distribution, relative to another.

Having reviewed the concepts of entropy and relative entropy, we define the \emph{initial privacy risk} as the KL divergence between the user's genuine profile $p$ and the population's tag distribution~$\bar{p}$, that is, $$\cR_0 = \oD(p\,\|\,\bar{p}).$$
Similarly, we define the \emph{(final) privacy risk}~$\mathcal{R}$ as the KL divergence between the user's apparent profile $t$ and the population's distribution, $$\cR=\oD(t\,\|\,\bar{p}).$$

Next, we justify the Shannon entropy and the KL divergence as measures of privacy when an attacker aims to individuate users based on their tag profiles.
The rationale behind the use of these two information-theoretic quantities as privacy metrics is documented in greater detail in~\cite{Parra13FGCS}.

Leveraging on a celebrated information-theoretic rationale by Jaynes~\cite{Jaynes82P}, the Shannon entropy of an apparent user profile may be regarded as a measure of privacy, or more accurately, anonymity.
The leading idea is that the method of types from information theory establishes an approximate monotonic relationship between the likelihood of a PMF in a stochastic system and its entropy. Loosely speaking and in our context, the higher the entropy of a profile, the more likely it is, and the more users behave according to it. Under this interpretation, entropy is a measure of anonymity, although not in the sense that the user's identity remains unknown. Entropy has, therefore, the meaning that the higher likelihood of an apparent profile can help the user go unnoticed. In fact, the apparent user profile makes the user more typical to an external observer, and hopefully, less attractive to an attacker whose objective is to target peculiar users.

If an aggregated histogram of the population is available as a reference profile, as we assume in this work, the extension of Jaynes' argument to relative entropy also gives an acceptable measure of anonymity.
The KL divergence is a measure of discrepancy between probability distributions, which includes Shannon's entropy as the particular case when the reference distribution is uniform.
Conceptually, a lower KL divergence hides discrepancies with respect to a reference profile, say the population's profile. Also, it exists a monotonic relationship between the likelihood of a distribution and its divergence with respect to the reference distribution of choice. This aspect enables us to deem KL divergence as a measure of anonymity in a sense entirely analogous to the above mentioned.

Under this interpretation, the KL divergence is, therefore, interpreted as an (inverse) indicator of the commonness of similar profiles in said population. As such, we should hasten to stress that the KL divergence is a measure of anonymity rather than privacy. The obfuscated information is the uniqueness of the profile behind the online activity, rather than the actual profile. Indeed, a profile of interests already matching the population's would not require perturbation.

\subsection{Privacy-Enhancing Techniques}
\label{PETs}
\noindent
Among a variety of PETs, this work focuses on those technologies that rely on the principle of \emph{tag forgery}. The key strengths of such tag-perturbation technique are its simplicity in terms of infrastructure requirements and its strong privacy guarantees, as users need not trust the social tagging platform, nor the network operator nor other peers.

In conceptual terms, tag forgery is a PET that may help users tagging online resources to protect their privacy. It consists of the simple idea that users may be willing to tag items that are unknown to them and that do not reflect their actual preferences, in order to appear as similar as possible to the average population profile. A simple example of such technique can be illustrated by thinking to a specific thematic community, such that of a group of individuals interested in jazz music. In this scenario if a user is particularly interested in rock music, their profile could be easily spotted and identified, as they would probably express interest towards artists and tracks that could be categorised outside of the jazz category.

When a user wishes to apply tag forgery, first they must specify a \emph{tag-forgery rate} $\rho \in [0,1]$. This rate represents the ratio of forged tags to total tags the user is disposed to submit.
Based on this parameter and exactly as in~\cite{Rebollo10IT}, we define the user's apparent tag profile as the convex combination $t = (1 - \rho)\,p + \rho\,r $.
Here $r$~is some \emph{forgery strategy} modeling the percentage of tags that the user should forge in each tag category.
Clearly, any forgery strategy must satisfy that $r_i \geqslant 0$ for all $i$ and that $\sum r_i = \rho$.

In this work, we consider three different forgery strategies, which result in three implementations of tag forgery, namely, optimised tag forgery~\cite{Rebollo10IT}, the popular TMN mechanism~\cite{Howe06B} and a uniform tag forgery.
The optimised tag forgery corresponds to choosing the strategy $r^*$ that minimises privacy risk for a given~$\rho$, that is,
$$r^* = \argmin_r \oD((1 - \rho)\,p + \rho\,r\,\|\,\bar{p}).$$
Please note that this formulation of optimised tag forgery relies on the appropriateness of the criteria optimised, which in turn depends on a number of factors. These are: the specific application scenario and the tag statistics of the users; the actual network and processing over-head incurred by introducing forged tags; the assumption that the tag-forgery rate $\rho$ is a faithful representation of the degradation in recommendation quality; the adversarial model and the mechanisms against privacy contemplated.

The TMN mechanism is described next. Said mechanism is a software implementation of query forgery developed as a Web browser add-on. It exploits the idea of generating false queries to a search engine in order to avoid user profiling from the latter. TMN is designed as a client-side software, specifically a browser add-on, independent from centralised infrastructure or third-party services for its operation. In the client software, a mechanism defined dynamic query lists has been implemented. Each instance of TMN is programmed to create an initial seed list of query terms that will be used to compute the first flow of decoys searches. The initial list of keywords is built from a set of RSS feeds from popular websites, mainly news sites, and it is combined with a second list of popular query words gathered from recently searched terms. When TMN is first enabled, and the user sends an actual search query, TMN intercepts the HTTP response returned from the search engine, and extracts suitable query-like terms that will be used to create the forged searches. Furthermore, the provided list of RSS feeds is queried randomly to substitute keywords in the list of seeds~\cite{a24}.

Because TMN sends arbitrary keywords as search queries, the user profile resulting from this forgery strategy is completely random~\cite{Chow09WPES}. Although the user possess the ability to add or remove RSS feeds that the extension will use to construct their bogus queries, there is no possible way to control which actual keywords are chosen. Moreover, the user has no control on the random keywords that are included in the bursts of bogus queries, since these are extracted from the HTTP response received from the actual searches that the user has performed. While TMN is a technique designed to forge \emph{search queries}, we have have implemented a TMN-like agent generating bogus \emph{tags}. To initialise our TMN-like agent we have considered an initial list of seed using RSS feeds from popular news sites, the sites included were the same ones that TMN uses in its built-in list of feeds. By querying the RSS feeds, a list of keywords was extracted. Hence, using the extracted keywords a distribution of tags into eleven categories was constructed, these eleven categories corresponds to the first taxonomy levels of the Open Directory Project (ODP) classification scheme~\cite{a22}. The profile obtained with this technique has then been assumed as a reference to implement a TMN agent and is denoted by the distribution~$w$.

Last but not least, the proposed uniform tag forgery strategy is constructed similarly to TMN. We have in fact supposed a TMN agent that would send disguise tags created according to a uniform distribution across all categories. More specifically, in the uniform forgery strategy we have that $r = u$. Table~\ref{ForgeryStrategies} summarises the tag-forgery strategies considered here.

\begin{table}
\caption{Summary of the tag-forgery strategies under study. In this work, we investigate three variations of a data-perturbative mechanism that consists of annotating false tags. The optimised tag forgery implementation corresponds to the strategy that minimises the privacy risk for a given forgery rate. The TMN-like approach generates false tags according to the popular privacy-preserving mechanism TrackMeNot~\cite{Howe06B}. The uniform approach considers the uniform distribution as forgery strategy.}
\label{ForgeryStrategies}
\centering
\renewcommand\arraystretch{1.4}
\begin{tabular}{c@{\hspace{25pt}}|c}
\multicolumn{2}{ c }{ } \\
Tag-forgery implementation & Forgery strategy $r$\\\hline
Optimised~\cite{Rebollo10IT} & $\argmin_r \oD ((1 - \rho)\,p + \rho\,r\,\|\,\bar{p})$\\
TMN~\cite{Howe06B} & $w$ (TMN distribution)\\
Uniform & $u$ (uniform distribution)\\\hline
\end{tabular}
\end{table}

\subsection{Similarity Metric}
\noindent
A recommender, or a recommendation system, can be described as an information filtering system that seeks to predict the rating or preference that a user would give to an item. For the purpose of our study, the idea of rating a resource or expressing a preference has been considered as the action of tagging an item. This assumption follows the idea that a user will most likely tag a resource if they happen to be interested in this resource.

In the field of recommendations systems, we may distinguish three main approaches to item recommendation: content-based, user-based and collaborative filtering~\cite{a25}.
In content-based filtering items are compared based on a measure of \emph{similarity}. The assumption behind this strategy is that items similar to those a user has already tagged in the past would be considered more relevant by the individual in question. If in fact a user has been tagging resources in certain categories with more frequency, it is more probable that they would also annotate items belonging to the same categories.

In user-based filtering, users are compared with other users based again on a defined measure of similarity. It is supposed, in this case, that if two or more users have similar interests, i.e. they have been expressing preference in resources in similar categories, items that are useful for one of them can also be significant for the others.

Collaborative filtering employs both a combination of the techniques described before as well as the collective actions of a group or network of users and their social relationships~\cite{collfilt}. In collaborative filtering then, not only the tags and categories that have been attached to a certain items are considered, but also what are called item-specific metadata are taken into account, these could be the item title or summary, or other content-related information~\cite{a04}.

In the coming sections, we shall use a generic content-based filtering algorithm~\cite{Lops11B} to evaluate the three variations of tag forgery described in Section~\ref{PETs}. 

We have chosen a content-based recommender because this class of algorithms models users and items as histograms of tags, which is essentially the model assumed for our adversary~\ref{mod profiles}. Loosely speaking a content-based recommendation system is composed of: a proper techniques for representing the items and users' profiles, a strategy to compare items and users and produce a recommendation. The field of content recommendation is particularly vast and developed in the literature and its applications are numerous. Recommendation systems in fact span different topics in computer science, information retrieval and artificial intelligence. 

For the scope of this job we are only concentrating on applying a suitable measure of similarity within items and users' profiles. The recommendation algorithm we have implemented therefore aims to find items that are closer to a particular user profile (i.e. more similar). Three commons measurement of similarity between objects are usually considered in the literature. These are namely: Euclidean distance, Pearson correlation and Cosine similarity~\cite{recc-sys-handbook}. 

The Euclidean distance is the simplest and most common example of a distance measure. The Pearson correlation is instead a measurement of the linear relationship between objects. While there are certainly different correlation coefficients that have been considered and applied, the Pearson correlation is among the most commonly used. 

Cosine similarity is another very common approach. It considers items as document vectors of an n-dimensional space and compute their similarity as the cosine of the angle that they form. We have applied this approach in our study.

More specifically, we have considered a cosine-based similarity~\cite{Markines09WWW} as a measure of distance between a user profile and an item profile.
The cosine metric is a simple and robust measure of similarity between vectors which is widely used in content-based recommenders.
Hence if $p_m = (p_{m,1},\ldots,p_{m,L})$ is the profile of user $u_m$ and $ q_n=(q_{n,1},\ldots,q_{n,L}) $  is the profile of item $i_n$, the cosine similarity between these two profiles is defined as
$$ s(p_m,q_n)= \frac { \sum_{l} {p_{m,l}}\, q_{n,l} } { \sqrt{\sum_l {p_{m,l}^2} } \sqrt{ \sum_l {q_{n,l}^2 } } }.$$

\subsection{Utility Metric}
\noindent
A utility metric is being introduced in order to evaluate the performances of the recommender and understand how these degrade with the application of a specific PET.
Prediction accuracy is among the most debated property in the literature regarding recommendation systems. For the purpose of this work it is assumed that a system providing on average more accurate recommendation of items would be preferred by the user. Furthermore the system is evaluated considering a content retrieval scenario where a user is provided with a ranked list of N recommended items, hence performances are evaluated in terms of ranking based metrics used in the Information Retrieval field of study~\cite{a16} .
The performance metric adopted is therefore among the most commonly used for ranked list prediction, i.e. precision at top V results. In the field of information retrieval, precision can be defined as the fraction of recommended items that are relevant for a target user~\cite{a17} . If the recommendation system evaluated retrieves V items, the previously defined ratio is precision at top V or P@V. Precision at top V is then a metric that measures how many relevant documents the user will find in the ranked list of results.
The overall performance value is then calculated by averaging the results over the set of all available users.
Considering a likely scenario, for which a user would be presented with a list of top-$V$ results that the system has considered most similar to their profile, we have evaluated precision of the recommender in two possible situations: with $V=30$ in one case and $V=50$ in the other.

\section{Architecture}
\label{arch}
\begin{figure}[ht!]
\centering
\includegraphics[scale=0.5]{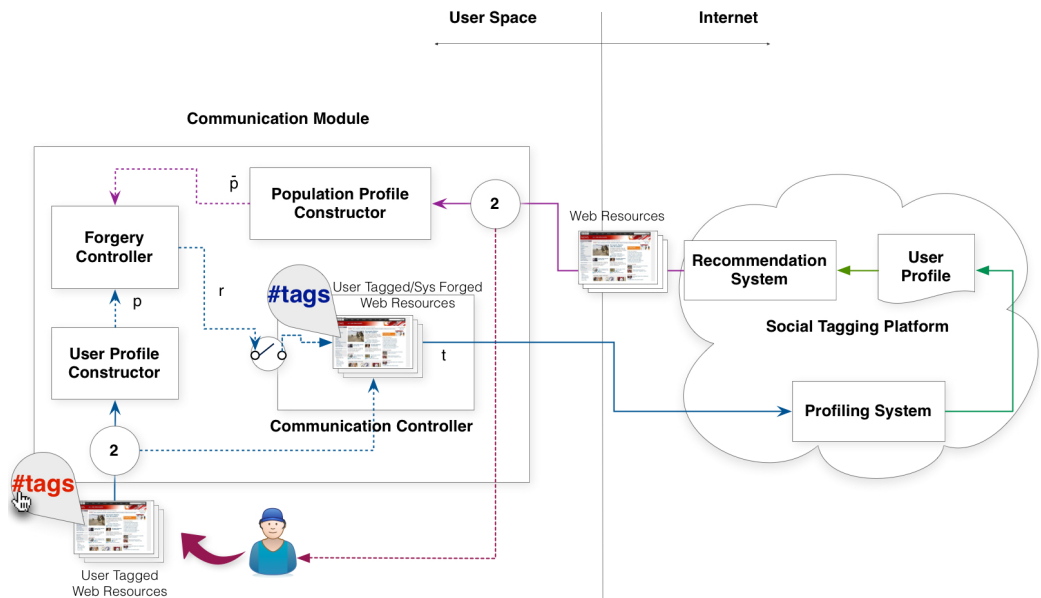}
\caption{The proposed architecture of a communication module managing the user data flows with a social tagging platform and implementing different possible forgery algorithms.}
\label{Arch}
\end{figure}

\noindent
In this section, we present an architecture of a communication module for the protection of user profiles in social tagging systems (Figure \ref{Arch}). We consider the case in which a user would retrieve items from a social tagging platform, and would occasionally submit annotations in the form of ratings or tags to the resource they would find interesting. This would be the case of a user browsing resources on StumbleUpon, tagging bookmarks on Delicious or exploring photos on Flickr. The social tagging platform would suggest web resources through its recommendation system that would gradually learn about the user interest, hence trying to suggest items more related to the user expressed preferences.

While the user would normally read the suggested documents, these would also be intercepted by the communication module, running as a software on the user space. This can be imagined as a browser extension analysing the communication between the user and the social tagging platform under consideration.

More generally, the communication module can be envisioned as a proxy or a firewall, i.e. a component between the user and the outside internet, responsible for filtering and managing the communication flows that the user generates. While the user would browse the internet the communication module would be in sleeping mode, and it could be turned on at the user's discretion only when visiting certain social tagging platforms. It is assumed that while the user would surf a certain platform, eventually annotating resources that they find relevant, they would receive and generate a stream of data, or more specifically a data flow. This is composed of the resources that the platform is sending to the user in the form of recommendation and of those that the user is sending back to the platform in the form of tagged items.

These data flows are analysed in the communication module by a component, the population profile constructor, and used to build a population profile of reference. We have supposed that these data streams would probably contain annotations that would help the module profiling the average population of users, together with other information regarding trends and current news. It is also possible that the module would contain specific, pre-compiled profiles, corresponding to particular population that the user would consider either safe or generic.

The user generated stream of data instead, composed by each annotated item, would be feeding the user profile constructor. This component would keep track of the actual expressed user preferences and feed this data into the forgery controller.

At this point the forgery controller would calculate a forgery strategy, that at the user discretion is either applied or not to the stream of tagged resources, and that would be sent to the social tagging platform, as the flow of data comprising the user activity. If the user kept the communication module on its off state, no forgery would modify the documents sent to the social tagging service, otherwise a certain stream of annotations would be computed and applied to certain resources.

This means that according to the strategy and a forgery rate that the user has chosen, the forgery controller would produce a number of bogus tags to certain items. These would be sent to the social tagging platform together with the actual user annotations. The user would hence present to the platform not their real profile, but an apparent profile $t$ resulting from both their real activity and the forged categorisation stream.

\subsection{Further considerations}
\label{further considerations}
\noindent

We would like to stress the fact that at the centre of our approach is the user. The communication module can in fact be used either to calculate a forgery strategy, or to simply warn the user when their privacy risk reaches a certain threshold. At this point the user would be presented with a possible forgery strategy and eventually are set of keywords and resources that could be used to produce bogus tags. We are aware that a mechanism generating tags could eventually produce a strategy introducing sensible topics in the user profile. We have, therefore, addressed this situation by using exclusively a curated list of websites and news portals whose content can be considered safe. In addition keywords in categories considered sensible could be excluded, either automatically or by the users. In our architecture is the user who ultimately decides whether to follow the recommendations proposed by our communication module or not. 

Additionally, it is worth mentioning that, if the user decided to reduce excessively the number of categories used to produce a possible forgery strategy, their user profile would inevitably exhibits a spike in activity in the chosen categories. As a consequence, the apparent user profile would probably become more identifiable to an external attacker. We therefore believe that although the user should be allowed to tweak their forgery strategy, they should also be informed of the consequences of applying some settings instead of others to the communication module.

We have also considered the possibility to implement our proposed architecture as a mobile application. We are aware this might add a computational, and networking overhead on the platform where the module will be installed, yet we also believe that in modern mobile platforms and personal computers this shall not be an issue. More importantly we believe that the benefit of controlling the user perceived profile shall overcome the cost of implementing the proposed architecture.

Profile data are in fact collected not only by social tagging platforms but also by websites, web applications and third parties even when the user is not connected to a personal account. Through tracking technologies and a networks of affiliated web sites users can be \emph{followed} online and their footprint collected for a variety of uses.
If aggregated, these data could reveal more over time that the same users initially intended. The data then turn from merely figures to piece of information able to describe users' identity and behaviours. Social engineering attacks could exploit users' profiles on different social networks to gather certain sensitive information. Similarly users' profiles crawling across different services and applications can disclose relevant facts about the users. It is, therefore, important for users to maintain a desired online privacy strategy. At the same time, this approach could also be implemented by developers and systems architects who need to be aware of the possible privacy and security implication of their work.

\section{Evaluation}
\label{eval}
\noindent
Evaluating how a recommender system would be affected when tag forgery is applied in a real world scenario is interesting for a different range of applications. We have particularly considered both the point of view of the privacy researcher interested in understanding how user privacy can be preserved, and also the perspective of an application developer willing to provide users with accurate recommendation regarding content and resources available on their platform.

\begin{figure}[htbp]
\centering
\includegraphics[scale=0.5]{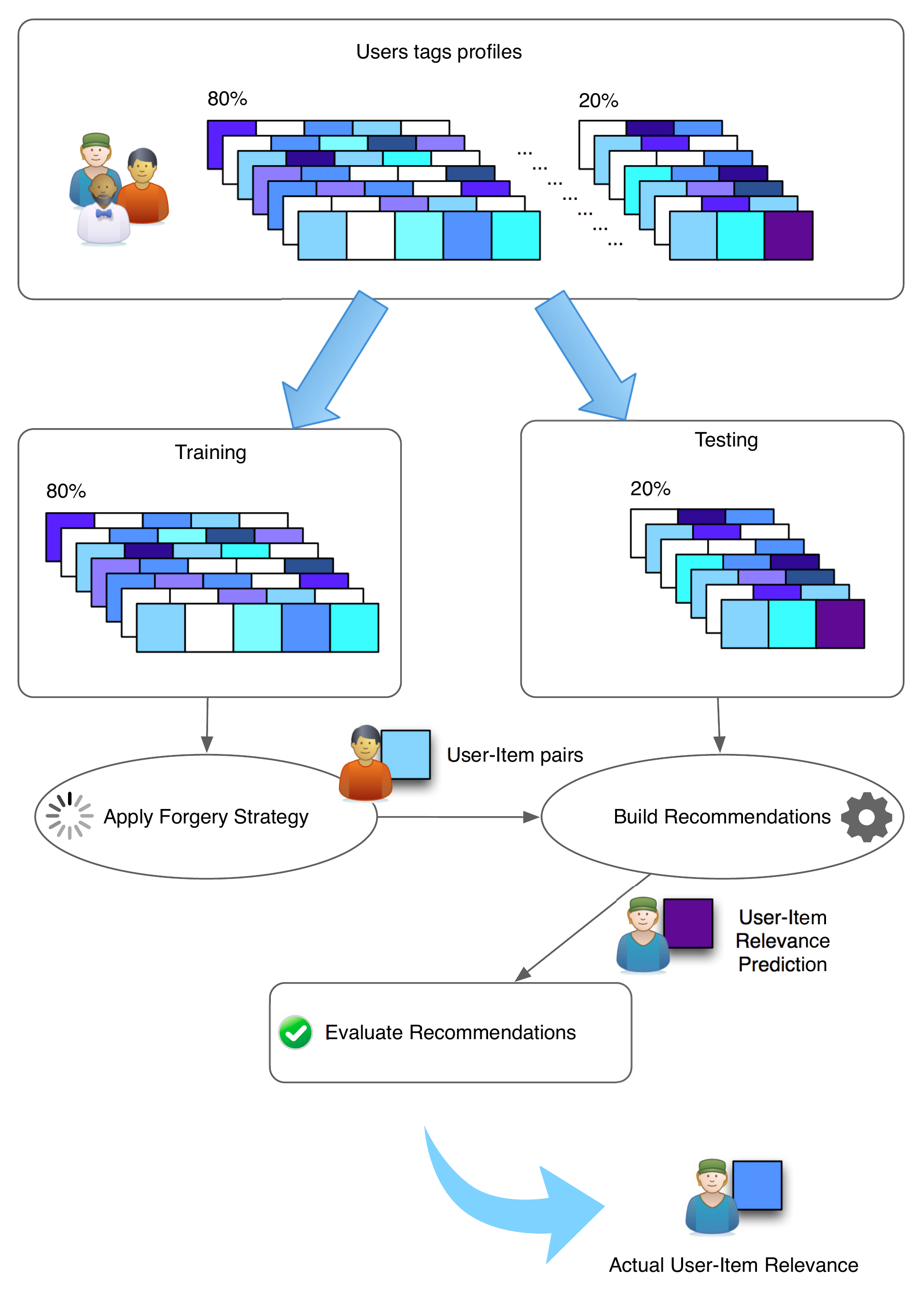}
\caption{Experimental methodology.}
\label{methodology}
\end{figure}

\begin{figure}[htbp]
\centering\hspace*{\fill}
\subfloat[Privacy risk $\mathcal{R}$ against forgery rate $\rho$ for all users.]%
{\includegraphics[scale=0.25]{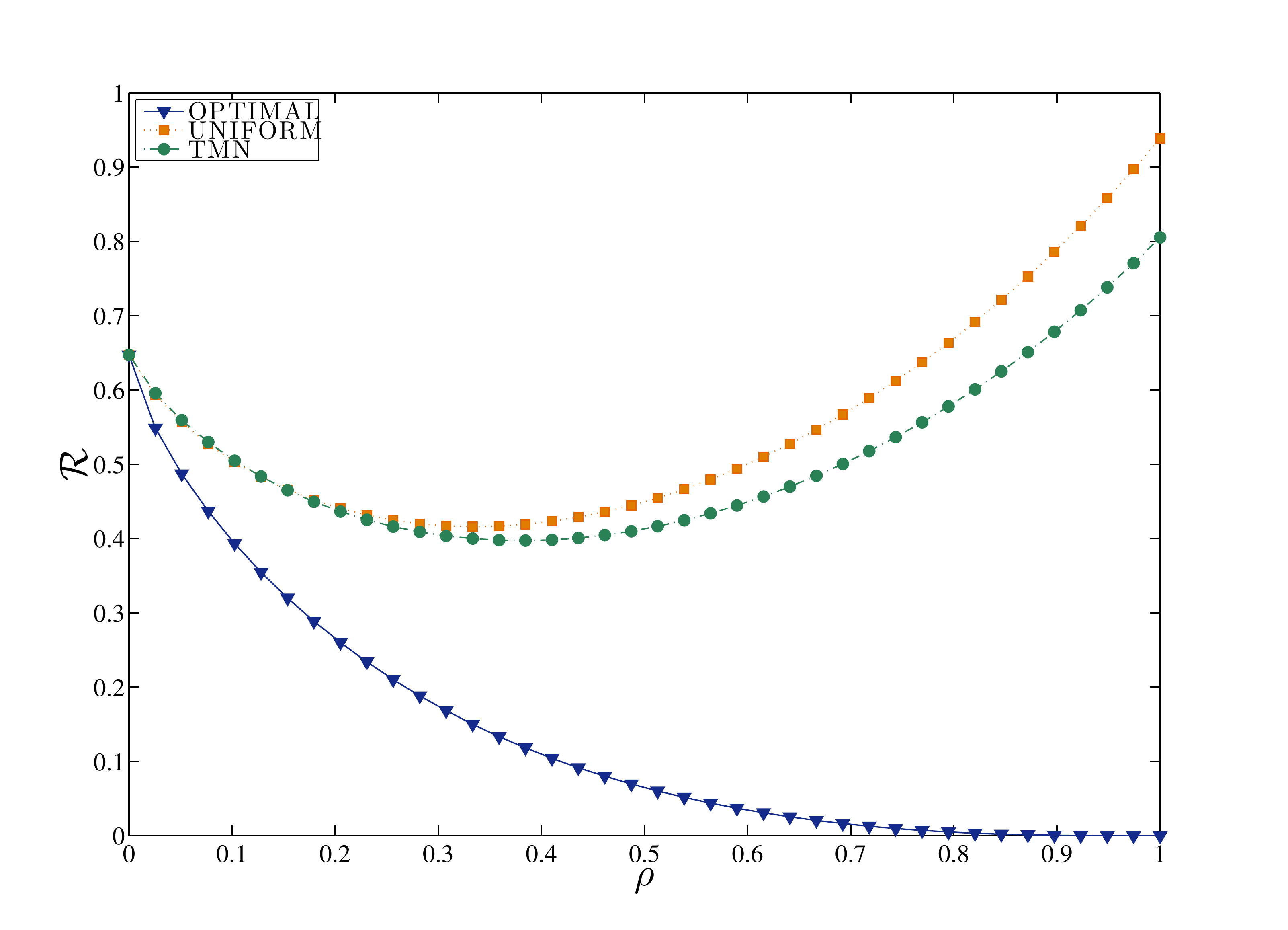}%
\label{Risk-Rho}}\hfill
\subfloat[Privacy risk $\mathcal{R}$ against forgery rate $\rho$ for a single user.]%
{\includegraphics[scale=0.25]{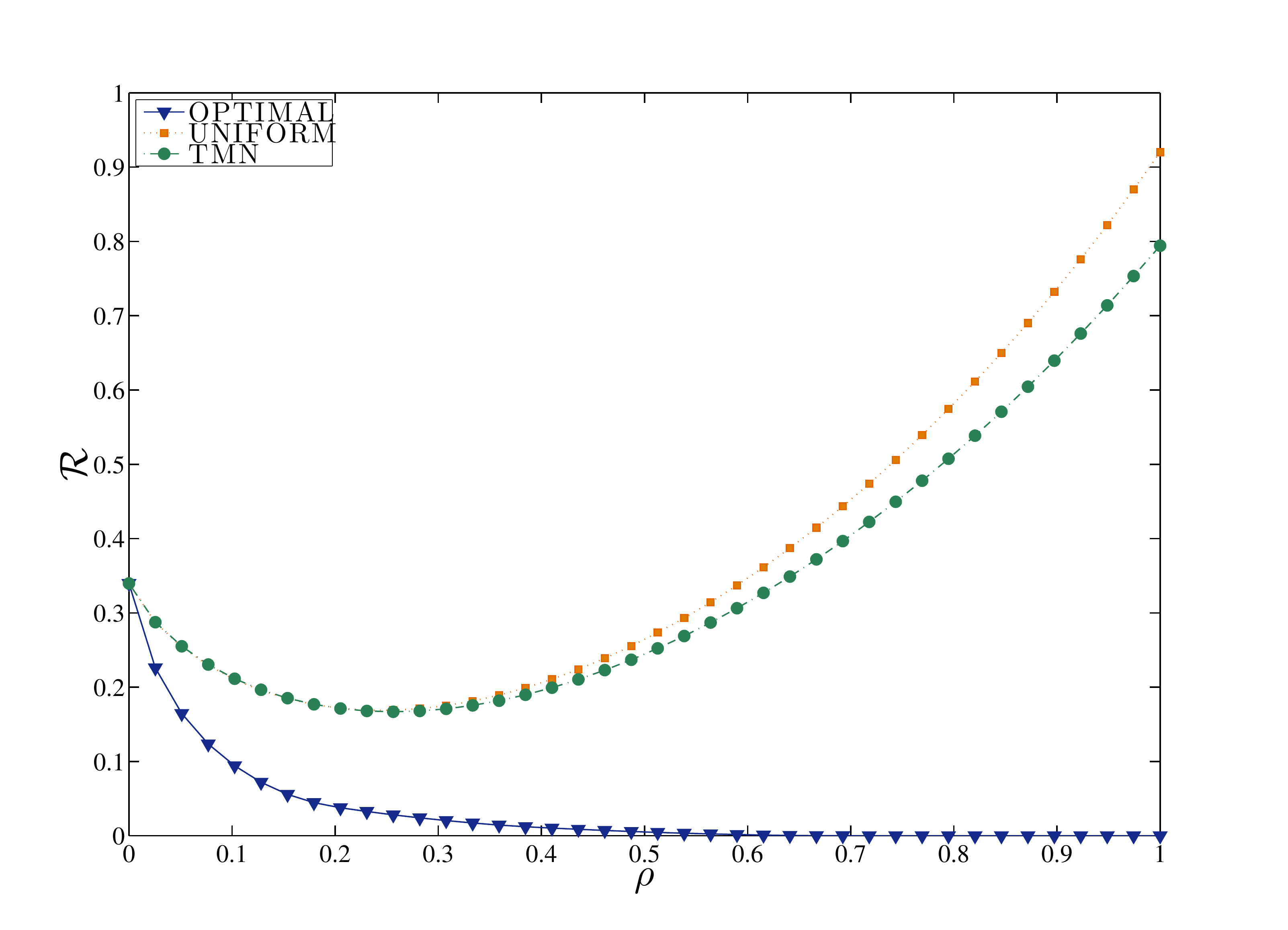}%
\label{Risk-Rho-Single}}\hspace*{\fill}
\caption[Privacy risk $\mathcal{R}$ against forgery rate $\rho$ for all users and for a single user applying a PET]%
{For the optimised forgery strategy the privacy risk $\mathcal{R}$ decreases with $\rho$. Naturally for $\rho = 0$ the privacy risk for all the users applying a technique is actually maximum, while it will approach 0\% when $\rho = 1 $. The graph shows how the optimised tag forgery strategy allows users to reduce more rapidly their privacy risk even for small values or $ \rho $. This confirms the intuitive assumption that applying a forgery strategy that actually modifies the user's apparent profile to increase its divergence from the average population profile, would produce the unfavourable result to make the user activity more easily recognised from a possible passive observer.}
\label{Privacy-Risk-Forgery-Ratio}
\end{figure}

\begin{table}[htbp]
\centering
\caption{Statistics regarding Delicious dataset}
\def\arraystretch{2.5}
\begin{tabular}{| l | r | l | r | }
\hline
\multicolumn{4}{|c|}{Statistics about the built dataset}            \\[2.5mm] \hline
Categories               & 11    & Users                   & 1867   \\[2.5mm] \hline
Item-Category Tuples     & 98998 & Avg. Tags per User      & 477.75 \\[2.5mm] \hline
Items                    & 69226 & Avg. Items per Category & 81044  \\[2.5mm]\hline
Avg. Categories per Item & 1.4   & Tags per item           & 13.06  \\[2.5mm]\hline
\end{tabular}
\label{tableDatasetStats}
\end{table}

\begin{figure}[htbp]
\centering\hspace*{\fill}
\subfloat[Average value of utility P@30 calculated for different values of $\rho$.]%
{\includegraphics[scale=0.25]{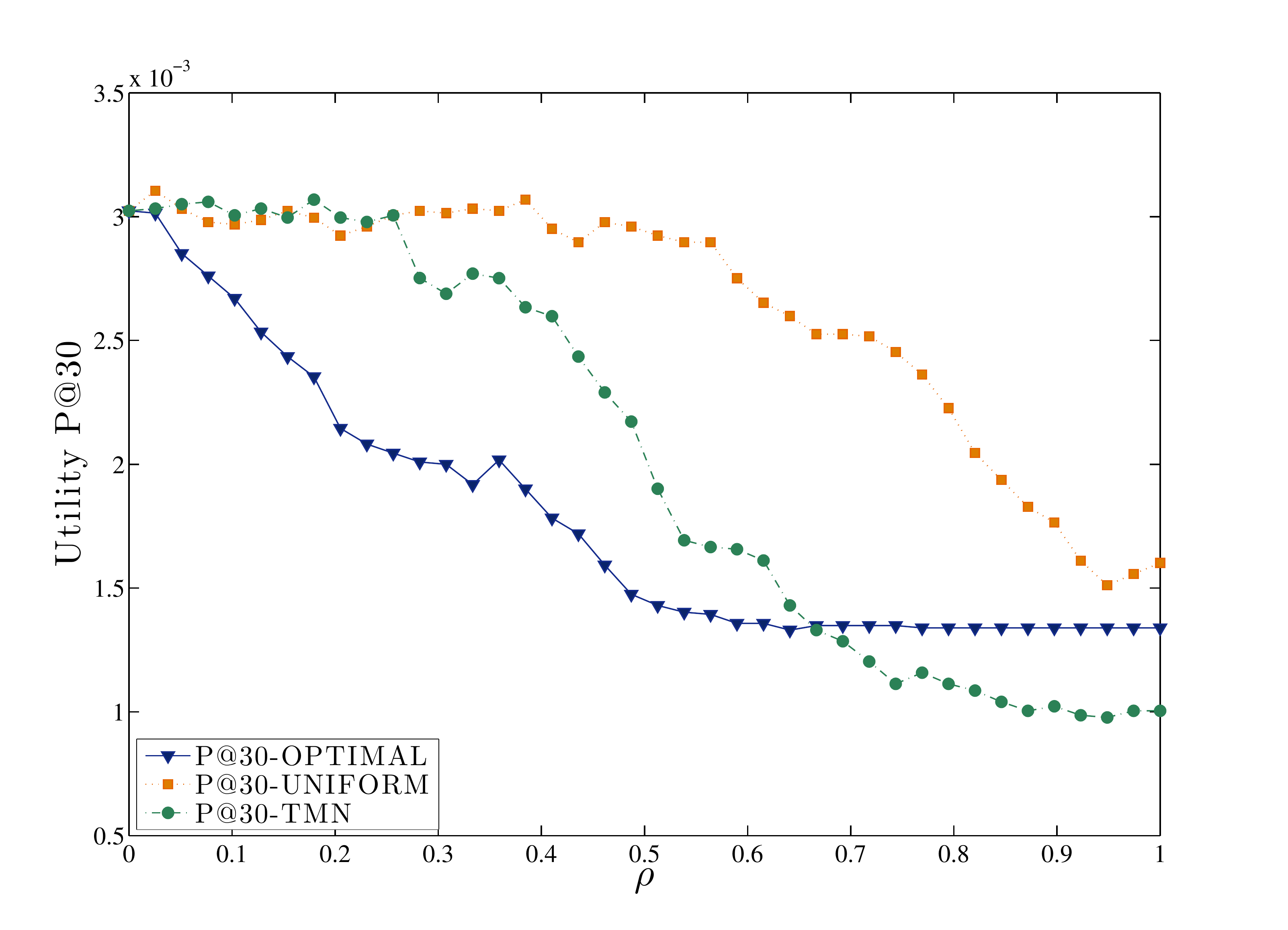}%
\label{P30-Rho}}\hfill
\subfloat[Average value of utility P@50 calculated for different values of $\rho$.]%
{\includegraphics[scale=0.25]{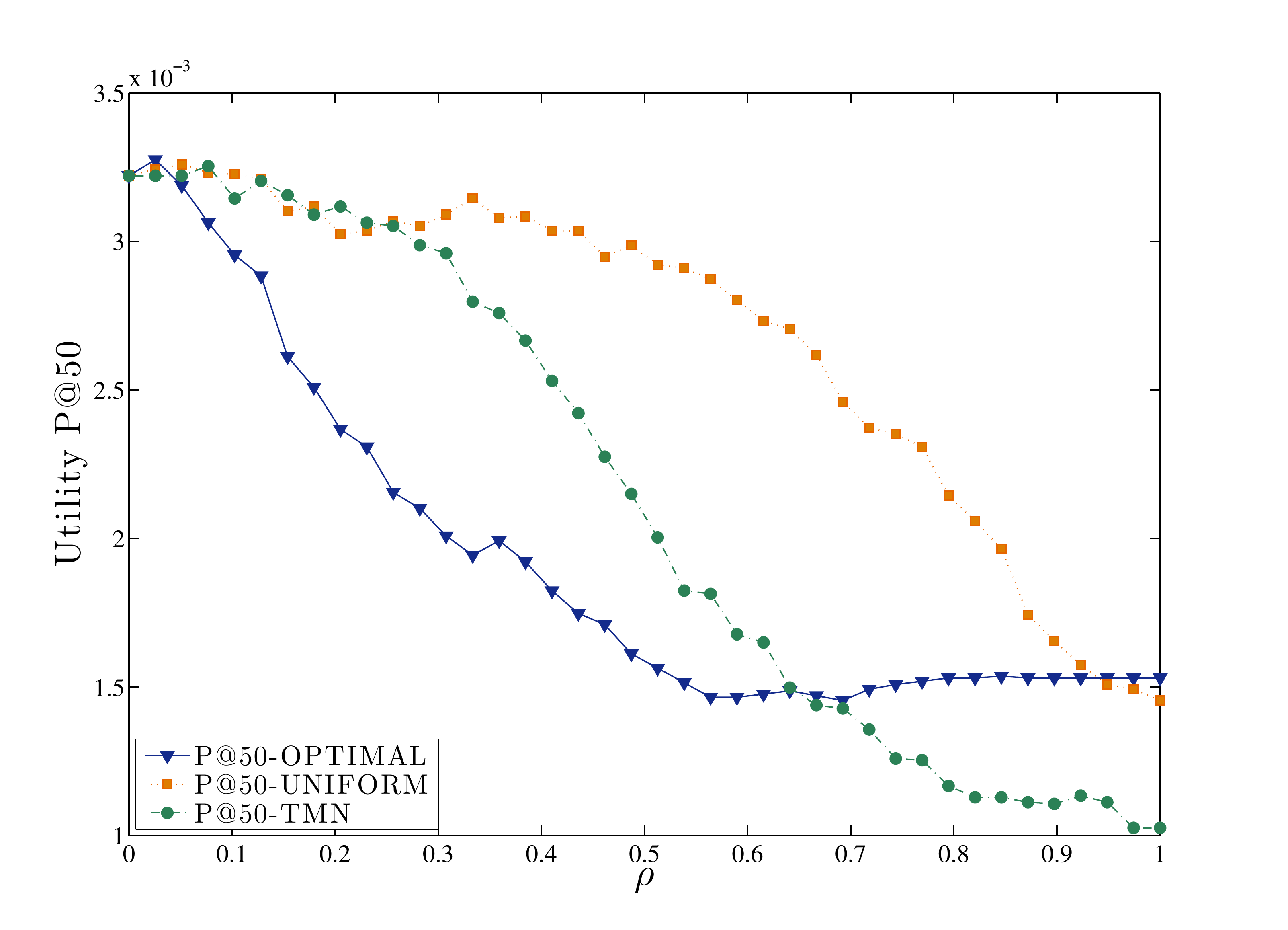}%
\label{P50-Rho}}\hspace*{\fill}
\caption[Average value of utility (P@30, P@50) calculated for different values of $\rho$.]%
{Average value of utility P@ calculated for different values of $\rho$, representing how applying a certain tag forgery strategy with a determined forgery rate affects the performance of a recommendation system, hence the user utility function. It is important to note that the measure of utility averaged across the user population is affected by statistical noise creating some glitches in the function that we can see attenuated if presenting each user with a larger list of results to choose from.}
\label{Utlity-Forgery-Ratio}
\end{figure}

\begin{figure}[htbp]
\centering\hspace*{\fill}
\subfloat[Privacy risk $\mathcal{R}$ against forgery rate $\rho$ for all users applying a PET considering only values of $\rho\leqslant 0.25$.]%
{\includegraphics[scale=0.25]{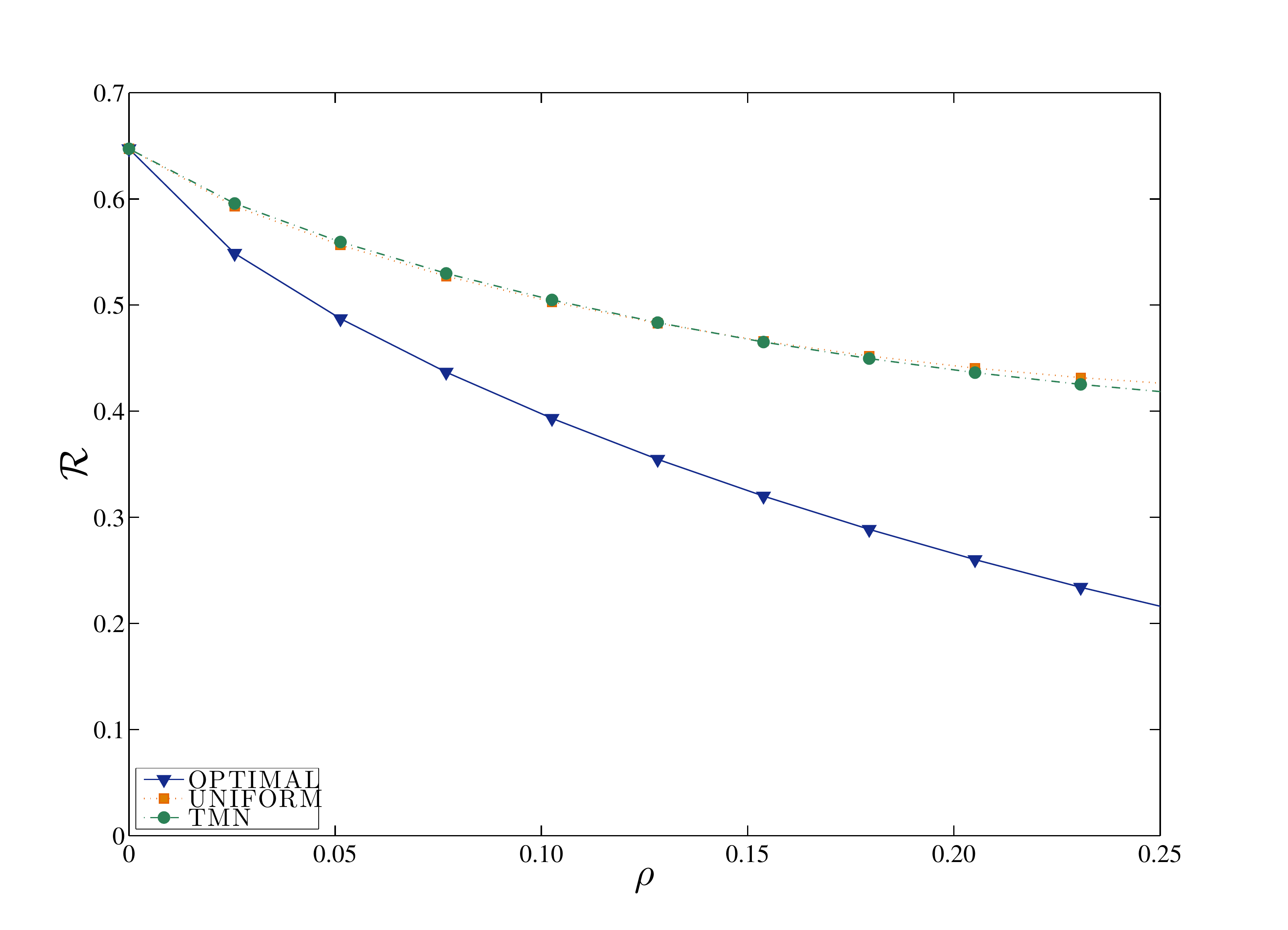}%
\label{Risk-Rho-Small}}\hfill
\subfloat[Average value of utility P@50 calculated for different values of $\rho\leqslant 0.25$.]%
{\includegraphics[scale=0.25]{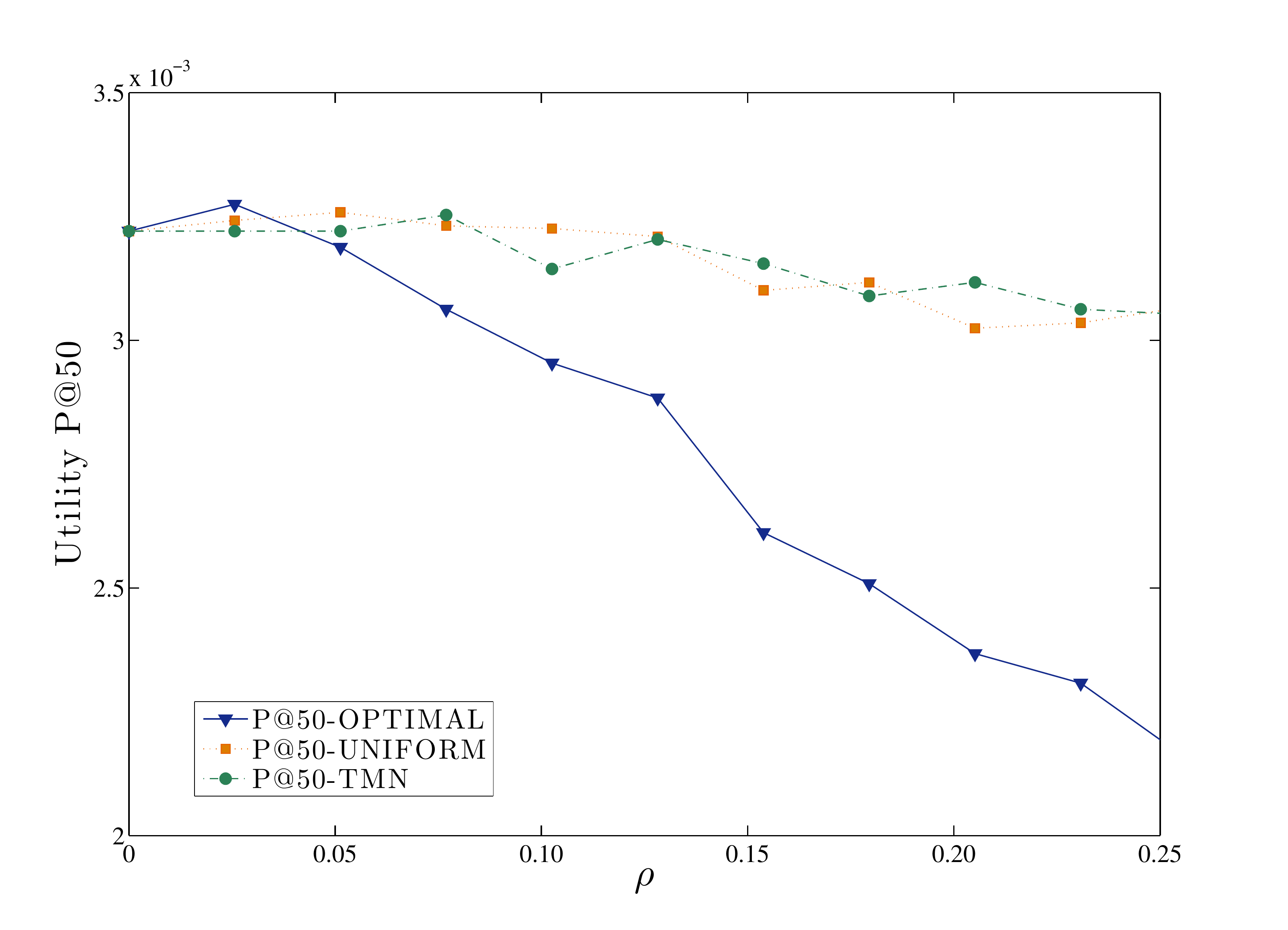}%
\label{Utility-Risk-Rho-Small}}\hspace*{\fill}
\caption[Privacy risk $\mathcal{R}$ against forgery rate $\rho$, compared with the average value of utility P@50, for small values of $\rho$, for all users applying a PET.]%
{It is interesting to note the ratio between the privacy risk $\mathcal{R}$ and the utility loss only for small values of $\rho$.}
\label{Privacy-Risk-Forgery-Ratio-Utlity-Small}
\end{figure}

\begin{figure}[tb]
\centering
\includegraphics[scale=0.5]{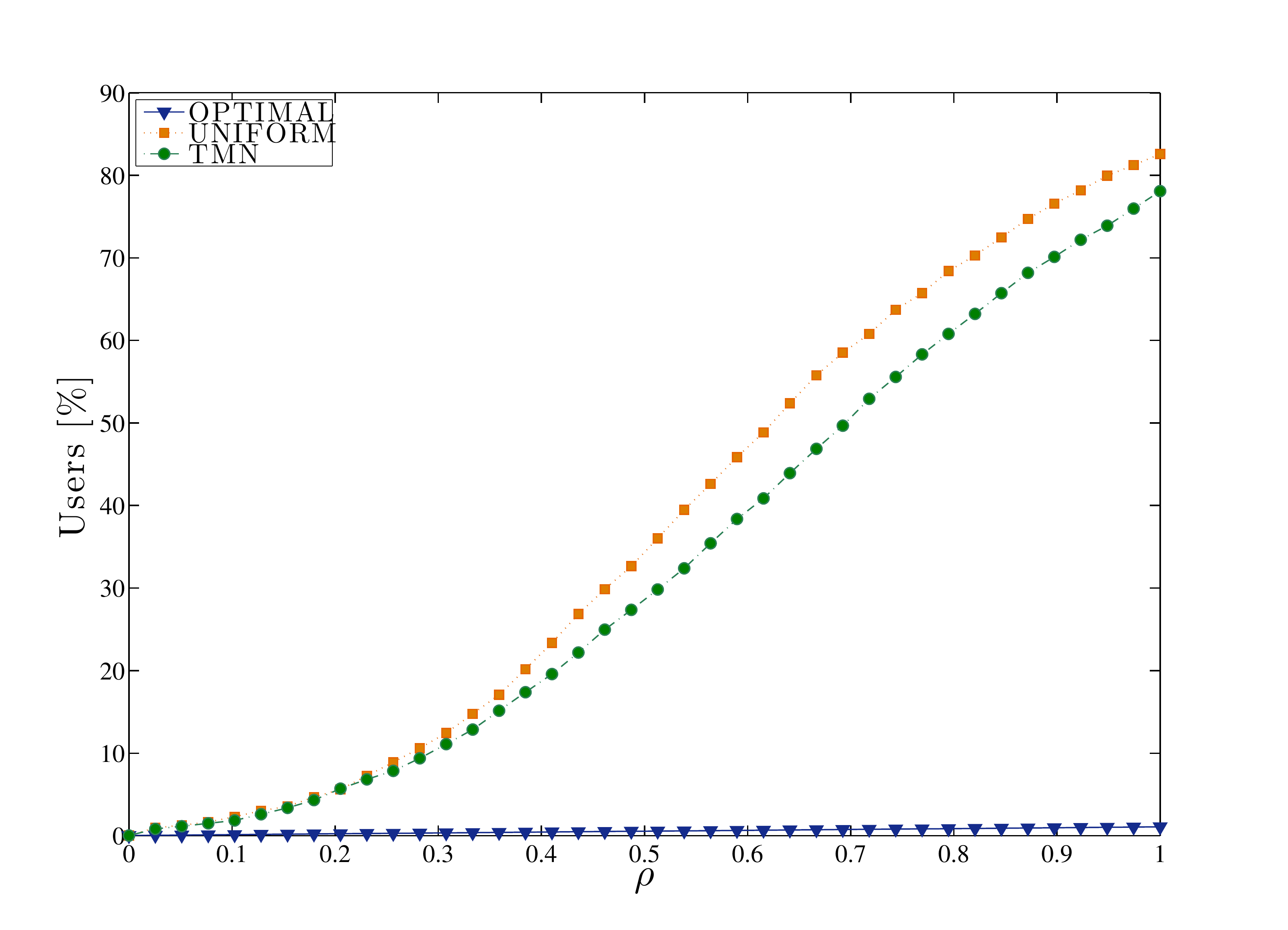}
\caption{Actual number of users increasing their privacy risk as a side effect of applying a certain strategy for a given value of $\rho$.}
\label{RhoIncr}
\end{figure}

Every PET must in fact ensure whether the semantic loss incurred in order to protect private data can be acceptable for practical use.

Thus, different tag forgery strategies were considered in a scenario where all the users were willing to apply the techniques. It was also considered that a user would try to apply a certain technique at different forgery rates, in order to evaluate how utility would be affected on average at each rate. When forgery rate is equal to zero it means the technique is not applied.

Hence, the overall utility for the recommender system, based on the applied forgery rate was evaluated against the privacy risk reduction calculated after each step.

In our simulated scenario, a user would ideally implement a possible PET at a time. We have therefore considered what percentage of utility the hypothetical user would lose when incrementing the ratio of forged tags with each strategy, consequently underlining what percentage of privacy risk reduction has gained in front of a certain loss in utility.

The user in this setup is presented over time with a list of top results, they would then decide to click or not on a number of these resources. This number divided by the total number of results gives us the percentage of items that the user has actually found interesting. Our utility metric is then evaluated considering the cases for which the user has been presented with the top 30 results, and the top 50 results.

Note that since in our experimental setting, we have split the data into a testing and a training set~\cite{a02,a03}, considering relevant only the items in the user's profile, it is not possible to evaluate items that are as yet unknown to the user but that could also be considered relevant (Figure \ref{methodology}).  In a real world application in fact, a user could be presented with results that are unknown to them, but that do reflect their expressed interestes. Therefore our estimation of precision is in fact an underestimation~\cite{a23}.

In order to evaluate the impact of a determined PET on the quality of the recommendation, and elaborate a study of the relationship between privacy and utility, a dataset rich in collaborative tagging information was needed.  Considering different social bookmarking platforms, Delicious was identified as a representative system. Delicious is a social bookmarking platform for web resources~\cite{a20}. The dataset containing Delicious data was obtained from the ones publicly available at the 2nd International Workshop on Information Heterogeneity and Fusion in Recommender Systems~\cite{a21}, accessible on http://ir.ii.uam.es/hetrec2011/datasets.html, and kindly hosted by  GroupLens research group at University of Minnesota. Furthermore, the dataset also contained category information about their items, this corresponds to the first and second taxonomy levels of the ODP classification scheme (Table ~\ref{tableDatasetStats})~\cite{a22}. The ODP project, now DMOZ, is the largest, most comprehensive human-edited directory of the Web, constructed and maintained by a passionate, global community of volunteers editors.

The chosen dataset specifically contains activity on the most popular tags in Delicious, the bookmarks tagged with those tags, and the users that tagged each bookmark. Starting from this specific set of users, the dataset also exhibits their contacts and contacts' contacts activity. Therefore it both covers a broad range of document?s topics while also presenting a dense social network ~\cite{a22}.

The experimental methodology is described also by Figure~\ref{methodology}. The dataset is randomly divided between two subsets, namely a testing and a training set. The training set contains 80\% of the items for each user, and was used to build the users' profiles. The testing set contained the remaining 20\% of the items tagged by each user, and was considered to evaluate (test) the recommender itself.

The first step of the experiment involved obtaining a metric of the recommender performance without applying any PET. The recommender would then produce estimation of how relevant an item potentially is for a user, by comparing the calculated user profile with each profile of the items in the testing set. This step would return a list of top items for each user. At this point our precision metric is calculated by verifying which of the top $V$ items have actually being tagged by each user. This process is repeated at each value of $\rho$ to understand how applying a different PET affects the prediction performances of a simple recommendation system. Please note that the three different PET have been considered independently for one another, i.e. the users would apply one of the techniques at a time and not a strategy involving a combination of the three.

\subsection{Experimental results}
\noindent
In our experimental setup, we have firstly evaluated what level of privacy users will reach implementing each of the strategies considered. Figure~\ref{Risk-Rho} shows how the application of the different PETs at different values of $\rho$ affect the privacy risk $\mathcal{R}$.

The first interesting result can be observed by considering how the privacy risk $\mathcal{R}$ is affected by the application of a certain PET. For values of $ \rho \in [0,0.25]$ (Figure~\ref{Risk-Rho-Small}), $\mathcal{R}$ is decreasing for all three strategies, although with optimised forgery this seem to be happening faster.

When larger values of $\rho$ are considered, the apparent user profile will most likely mimic the profile of either the population distribution, in the case of optimised forgery, the TMN distribution in the case of TMN and the uniform distribution in the case of uniform forgery.
If we consider this apparent effect, we understand why, while the privacy risk approaches 0 in the case of optimised forgery, it actually increases both for TMN and uniform forgery (Figure~\ref{Risk-Rho}). Recalling that our privacy metric, and adversary model, consider the case for which a possible attacker would try to isolate a certain user from the rest of the population, applying a forgery strategy that would generate an apparent profile $t$ that would increase the divergence from an average profile, would actually result in making the user more easily identified from a possible observer.

This undesirable consequence is also more eloquently present when applying the uniform strategy, in fact as the user apparent profile approached the uniform distribution for higher values of $\rho$, it would become evident to an external observer which users are forging their tags according to this strategy.

In the case of optimised forgery instead, privacy risk decreases with $\rho$. Naturally for $\rho = 0$ the privacy risk for all the users applying a technique is actually maximum, while it will approach 0\% when $\rho = 1 $. It is particularly interesting to see how our optimised tag forgery strategy allows users to reduce their privacy risk  more rapidly even for small values or $ \rho $.

We have therefore measured the total number of users that would actually increase their privacy risk as a consequence of having applied a certain PET (Figures~\ref{RhoIncr}). It is surprisingly striking to observe how almost 90\% of the total number of users, when applying TMN or uniform forgery, would make their apparent profile more recognisable than without implementing any PET.
This reflects the intuitive assumption that in order to conceal the actual user's profile, with the privacy metric considered throughout this work, it would be advisable to make it as close as possible to an average profile of reference, so that it is not possible to individuate it, or in other words to distinguish it from the average population profile.

We then have evaluated how our utility metric was affected by the application of the tag forgery strategies, for different values of $\rho$. We have considered two situations to evaluate our utility metric. In the first case the user would be presented with the top 30 results, and in the second with the top 50. This allowed us, not only to evaluate the impact of noise on the metric itself, but also to consider the impact of a certain strategy over longer series of results.

Figure~\ref{P50-Rho} and Figure~\ref{P30-Rho}, shows the obtained utility versus the rate of tag forgery applied, this has been evaluated again for optimised forgery, uniform forgery, and TMN strategy, in order to understand how these PETs perform in the described scenario.

In this case we noticed how a uniform forgery strategy, which generates bogus tags according to a uniform distribution across all categories, is able to better preserve utility than either optimised tag forgery or TMN, especially for bigger forgery ratios.

What we found particularly relevant in our study is that for smaller values of $\rho$, hence for a forgery rate up to 0.1, corresponding to a user forging 10\% of their tags, our optimised forgery strategy shows a privacy risk reduction $\mathcal{R}$  of almost 34\% opposed to a degradation in utility of 8\%. This result is particularly representative of the intuition that it is possible to obtain a considerable increase in privacy, with a modest degradation of performance of the recommender system, or in other words a limited utility loss (Figure~\ref{Utility-Risk-Rho-Small}).

The results obtained therefore present a scenario where applying a tag forgery technique perturbs the profile observed from the outside, thus enabling users to protect their privacy, in exchange of a small semantic loss if compared to the privacy risk reduction. The performance degradation measured for the recommendation systems, is small if compared to the privacy risk reduction obtained by the user when applying the forgery strategy considered.

\section{Conclusions}
\label{conclusions}
\noindent
Information filtering systems that have been developed to predict users' preferences, and eventually use the resulting predictions for different services, depend on users revealing their personal preferences by annotating items that are relevant to them.  At the same time, by revealing their preferences online users are exposed to possible privacy attacks and all sorts of profiling activities by legitimate and less legitimate entities.

Query forgery arises, among different possible PETs, as a simple strategy in terms of infrastructure requirements, as no third parties or external entities need to be trusted by the user in order to be implemented.

However, query forgery poses a trade-off between privacy and utility. Measuring utility by computing the list of useful results that a user would receive from a recommendation system, we have evaluated how three possible tag forgery techniques would perform in a social tag application. With this in mind a dataset for a real world application, rich in collaborative tagging information has been considered.

Delicious provided a playground to calculate how the performance of a recommendation system would be affected if all the users implemented a tag forgery strategy. We have hence considered an adversary model where a passive privacy attacker is trying to profile a certain user. The user in response, adopts a privacy strategy aiming at concealing their actual preferences, minimising the divergence with the average population profile. The results presented show a compelling outcome regarding  how implementing different PETs can affect both user privacy risk, as well as the overall recommendation utility.

We have firstly observed how while the privacy risk $\mathcal{R}$  decreases initially, for smaller values of $\rho$ (for both TMN and uniform forgery), it increases as bigger forgery ratios are considered. This is because the implied techniques actually modify the apparent user profile to increase its divergence from the average population profile. This actually makes the user activity more easily recognised from a possible passive observer. On the other hand, optimised forgery has been designed to minimise the divergence between the user and the population profile, therefore the effect described is not observed in this case.

Considering this unfavourable effect, we have computed the number of users that would actually increase their privacy risk. This particular result showed how applying a certain PET could actually be detrimental to the user's privacy: if the user implemented a strategy that is not accurately chosen, they would be exposed to a higher privacy risk than the one measured before applying the PET.
Observing how the application of a PET affects utility, we have found out that especially for a small forgery rate (up to 20\%) it is possible to obtain a consistent increase in privacy, or privacy risk reduction, against a small degradation of utility. This reflects the intuition that users would be able to receive personalised services while also being able to reasonably protect their privacy and their profiles from possible attackers.

This study furthermore shows in a simple experimental evaluation, of a real world application scenario, how the performances degradation of a recommendation system, is small if compared to the privacy risk reduction offered by the application of these techniques. This opens many possibilities and paths that need to be explored to better understand the relationship between privacy and utility in recommendation systems. In particular it would be interesting to explore other definitions of the metrics proposed and apply these on different class of recommendation systems.

As future research lines, we shall investigate how other information filtering models are affected by the application of certain PET. Specifically we shall consider researching how different aspects of users' activities are categorised and profiled by information filtering systems, and what counter-measures can be taken to protect user privacy.

\section*{Acknowledgment}
The authors would like to thank Iv\'an Cantador, with the Department of Computer Science at the Universidad Aut\'onoma de Madrid, for providing us with the tag categorisation of the data set.
This work was partly supported by the Spanish Government through projects CONSEQUENCE (TEC2010-20572-C02-02) and EMRISCO (TEC2013-47665-C4-1-R). D. Rebollo-Monedero is the recipient of a Juan de la Cierva postdoctoral fellowship, JCI-2009-05259, from the Spanish Ministry of Science and Innovation.

\section*{References}
\bibliographystyle{elsarticle-num}
\bibliography{Bibliography/StringAbbreviated,Bibliography/Security,Bibliography/InfoTheory,Bibliography/LosslessCoding,Bibliography/LossyCoding,Bibliography/MathStatSigPro,Bibliography/Classification,Bibliography/Applications,Bibliography/ReferencesTRIPP,Bibliography/SemanticWeb,Bibliography/InsubriaReferences,Bibliography/rfc,Bibliography/Silvia_bibliography}
\end{document}